\documentclass[12pt,a4paper]{article}

\usepackage{amsmath}
\usepackage{amssymb}
\usepackage{epsfig}
\usepackage{graphicx}
\usepackage{cite}

\newcommand{\capdef}{}
\newcommand{\mycaption}[2][\capdef]{\renewcommand{\capdef}{#2}%
       \caption[#1]{{\footnotesize #2}}}
\makeatletter
\renewcommand{\fnum@table}{\textbf{\tablename~\thetable}}
\renewcommand{\fnum@figure}{\textbf{\figurename~\thefigure}}
\makeatother

\newcounter{myenumi}

\renewcommand{\themyenumi}{\roman{myenumi}}
{\end{list}}

\newlength{\myem}
\newcounter{mysubequation}[equation]

\makeatletter
\renewcommand{\section}{\@startsection{section}{1}{0em}{-\baselineskip}%
{\baselineskip}{\normalfont\large\bfseries}}
\renewcommand{\subsection}%
{\@startsection{subsection}{2}{0em}{-0.7\baselineskip}%
{0.7\baselineskip}{\normalfont\bfseries}}
\makeatother

\newcommand{\eVq}  {\text{eV}^2}
\newcommand{\stheta}{\sin^22\theta_{13}}
\newcommand{\deltacp}{\delta_\mathrm{CP}}

\newcommand{\minos}{MINOS}

\newcommand{\icarus}{ICARUS}
\newcommand{\opera}{OPERA}
\newcommand{\ttk}{T2K}
\newcommand{\nova}{NO$\nu$A}
\newcommand{\dchooz}{D-Chooz}
\newcommand{\RII}{Reactor-II}

\begin{document}

\renewcommand{\thefootnote}{\alph{footnote}}

\begin{flushright}
SISSA 79/2005/EP\\
\end{flushright}

\vspace*{1cm}

\renewcommand{\thefootnote}{\fnsymbol{footnote}}

\begin{center} 
\Large\textbf{Neutrino oscillations: Current status and
prospects}\footnote{Talk given at the XXIX International Conference of
Theoretical Physics, ``Matter To The Deepest: Recent Developments In
Physics of Fundamental Interactions'', 8--14 September 2005, Ustron,
Poland.}
\end{center}
\renewcommand{\thefootnote}{\it\alph{footnote}}

\vspace*{.8cm}

\begin{center} 
{\bf Thomas Schwetz}
\end{center}
{\it
\begin{center}
       Scuola Internazionale Superiore di Studi Avanzati (SISSA)\\
       Via Beirut 2--4, I--34014 Trieste, Italy
\end{center}}

\vspace*{0.5cm}

\begin{abstract}
   I summarize the status of neutrino oscillations from world neutrino
   oscillation data with date of October 2005. The results of a global
   analysis within the three-flavour framework are
   presented. Furthermore, a prospect on where we could stand in
   neutrino oscillations in ten years from now is given, based on a
   simulation of upcoming long-baseline accelerator and reactor
   experiments.
\end{abstract}

\renewcommand{\thefootnote}{\arabic{footnote}}
\setcounter{footnote}{0}

\section{Introduction}

In the last ten years or so we have witnessed huge progress in neutrino
oscillation physics. The outstanding experimental results lead to
quite a clear overall picture of the neutrino sector. We know that
there are two mass-squared differences separated roughly by a factor
of 30, and in the lepton mixing matrix there are two large mixing
angles, and one mixing angle which has to be small.
In the first part of this talk I review the present status of neutrino
oscillations by reporting the results of a global analysis of latest
world neutrino oscillation data from solar, atmospheric, reactor and
accelerator experiments. This analysis is performed in the
three-flavour framework and represents an update of the work published
in Refs.~\cite{Maltoni:2003da,Maltoni:2004ei}.

The recent developments in neutrino oscillations triggered a lot of
activity in the community, and many new neutrino oscillation
experiments are under construction, or under active investigation, to
address important open questions, such as the value of the small mixing
angle $\theta_{13}$, leptonic CP violation and the type of the
neutrino mass hierarchy. In the second part of the talk I try to give
an outlook, where we could stand in about ten years from now. These
results are based on a simulation of up-coming long-baseline
accelerator and reactor experiments, which are expected to deliver
physics results within the anticipated time
scale~\cite{Huber:2003pm,Huber:2004ug}.

Three-flavour neutrino oscillations are described in general by the two
independent mass-squared differences $\Delta m^2_{21}$, $\Delta
m^2_{31}$, three mixing angles $\theta_{12},\theta_{23},\theta_{13}$, and
one complex phase $\deltacp$. Throughout this work I will use the
standard parameterization for the PMNS lepton mixing matrix
\begin{equation}
U =  
  \left(\begin{array}{ccc} 
    1&0&0\\ 0 & c_{23} & s_{23}\\ 0 & -s_{23} & c_{23}
  \end{array}\right) 
  \left(\begin{array}{ccc}
    c_{3}&0 & e^{-i\deltacp}s_{13} \\ 0 & 1 & 0 \\ 
    -e^{i\deltacp} s_{13}&0 & c_{13} 
  \end{array}\right)
  \left(\begin{array}{ccc}
    c_{12} & s_{12}& 0\\ -s_{12} & c_{12}&0\\0&0&1 
  \end{array}\right)
\end{equation}
with the abbreviations $s_{jk} \equiv \sin\theta_{jk}$, $c_{jk} \equiv
\cos\theta_{jk}$. The type of the neutrino mass hierarchy is
determined by the sign of $\Delta m^2_{31}$: $\Delta m^2_{31} > 0$
corresponds to the normal hierarchy and $\Delta m^2_{31} < 0$ to the
inverted one.


\section{Present status of three-flavour neutrino oscillations}

In Tab.~\ref{tab:3nu-summary} I summarize the present status of
three-flavour neutrino oscillation parameters. The numbers are
obtained from a global analysis of current oscillation data from
solar~\cite{solar,ahmad:2002ka,sno2005}, atmospheric~\cite{atm},
reactor~\cite{kamland,chooz}, and accelerator~\cite{k2k} data. Details
of the analysis can be found in Ref.~\cite{Maltoni:2004ei} and
references therein. In the following I give some brief comments on the
determination of the 'atmospheric' and the 'solar' parameters, and on
the bound on $\theta_{13}$.
 
\begin{table}
\centering
\begin{tabular}{|@{\quad}l@{\quad}|@{\quad}c@{\quad}|@{\quad}c@{\quad}|@{\quad}c@{\quad}|}
   \hline\hline
   parameter & bf$\pm 1\sigma$ & $1\sigma$ acc. & 3$\sigma$ range\\
   \hline
   \rule[-2mm]{0ex}{3ex}
   $\Delta m^2_{21}  \: [10^{-5}\eVq]$ & $7.9\pm 0.3$ & 4\% & $7.1-8.9$ \\
   \rule[-2mm]{0ex}{3ex}
   $|\Delta m^2_{31}|\: [10^{-3}\eVq]$ & $2.2^{+0.37}_{-0.27}$ & 14\% & $1.4-3.3$ \\
   \hline
   \rule[-2mm]{0ex}{4ex}
  $\sin^2\theta_{12}$ & $0.31^{+0.02}_{-0.03}$ &  9\% & $0.24-0.40$ \\
   \rule[-2mm]{0ex}{3ex}
  $\sin^2\theta_{23}$ & $0.50^{+0.06}_{-0.05}$ & 11\% & $0.34-0.68$ \\
   \rule[-2mm]{0ex}{3ex}
  $\sin^2\theta_{13}$ & $-$  &$-$  & $\leq 0.046$ \\
  \hline\hline
\end{tabular}
\mycaption{Best fit values (bf), $1\sigma$ errors, relative accuracies
  at $1\sigma$, and $3\sigma$ allowed ranges of three-flavour neutrino
  oscillation parameters from a combined analysis of global
  data, updated from Ref.~\cite{Maltoni:2004ei}.}
  \label{tab:3nu-summary}
\end{table}

\begin{figure}
\centering
\includegraphics[width=0.55\textwidth]{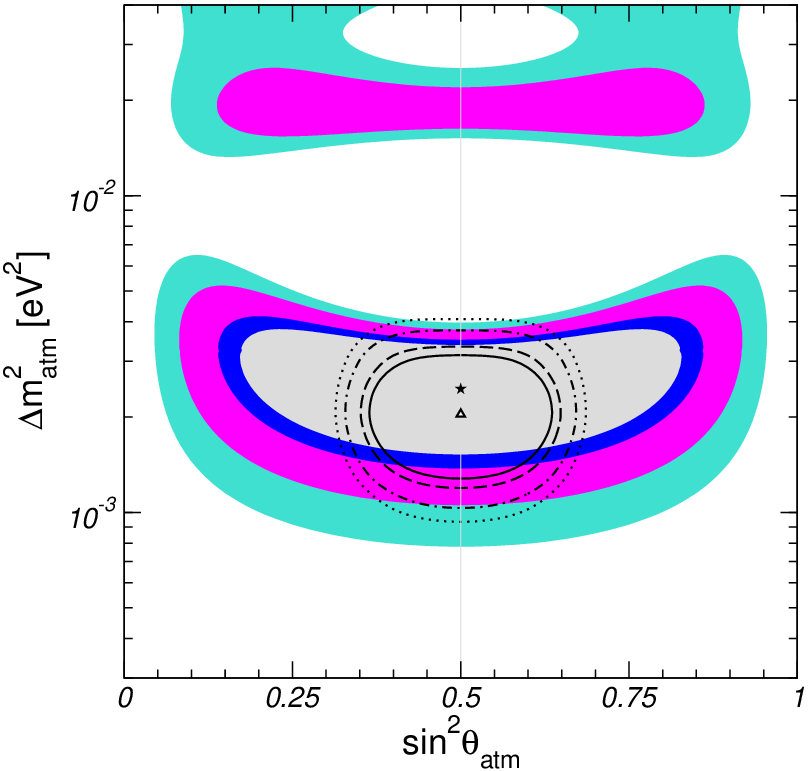}%
\mycaption{Allowed regions for $\sin^2\theta_{23}$ and $\Delta m^2_{31}$
  at 90\%, 95\%, 99\%, and $3\sigma$ C.L.\ for atmospheric neutrino data
  (contour lines) and the K2K long-baseline experiment (coloured
  regions).}
  \label{fig:atm}
\end{figure}

{\it The 'atmospheric parameters'.} In Fig.~\ref{fig:atm} I show the
allowed regions for $\theta_{23}$ and $\Delta m^2_{31}$ from separate
analyses of Super-K atmospheric neutrino data~\cite{atm}, and data
from the K2K long-baseline experiment~\cite{k2k}. The latter probes
the $\nu_\mu$ disappearance oscillation channel in the same region of
$\Delta m^2$ as explored by atmospheric neutrinos. The neutrino beam
is produced at the KEK proton synchrotron, and originally consists of
98\% muon neutrinos with a mean energy of 1.3~GeV. The $\nu_\mu$
content of the beam is observed at the Super-K detector at a distance
of 250~km, where 107 events have been detected, whereas
$151^{+12}_{-10}$ have been expected for no
oscillations. Fig.~\ref{fig:atm} illustrates that the neutrino
mass-squared difference indicated by the $\nu_\mu$ disappearance
observed in K2K is in perfect agreement with atmospheric neutrino
oscillations. Hence, K2K data provide the first confirmation of
oscillations with $\Delta m^2_{31}$ from a man-made neutrino
source. K2K gives a rather weak constraint on the mixing angle due to
low statistics in the current data sample, and the constraints on
$\sin^2\theta_{23}$ of Tab.~\ref{tab:3nu-summary} are dominated
by atmospheric data. Both data sets give a best fit point of
$\theta_{23} = \pi/4$, {\it i.e.}\ maximal mixing.

In our analysis of atmospheric neutrino data we neglect the small
contribution of oscillations with $\Delta m^2_{21}$.  Taking into
account this sub-leading effect, in
Refs.~\cite{Gonzalez-Garcia:2004cu,Fogli:2005cq} a small deviation
from maximal mixing was found, due to an excess of sub-GeV $e$-like
events. This indication currently is not statistically significant
(about $0.5\sigma$), and so-far it has not been confirmed by a
three-flavour analysis of the Super-K collaboration~\cite{kajita}.

\begin{figure}
\centering
\begin{tabular}{c@{\quad}c}
\includegraphics[height=7.5cm]{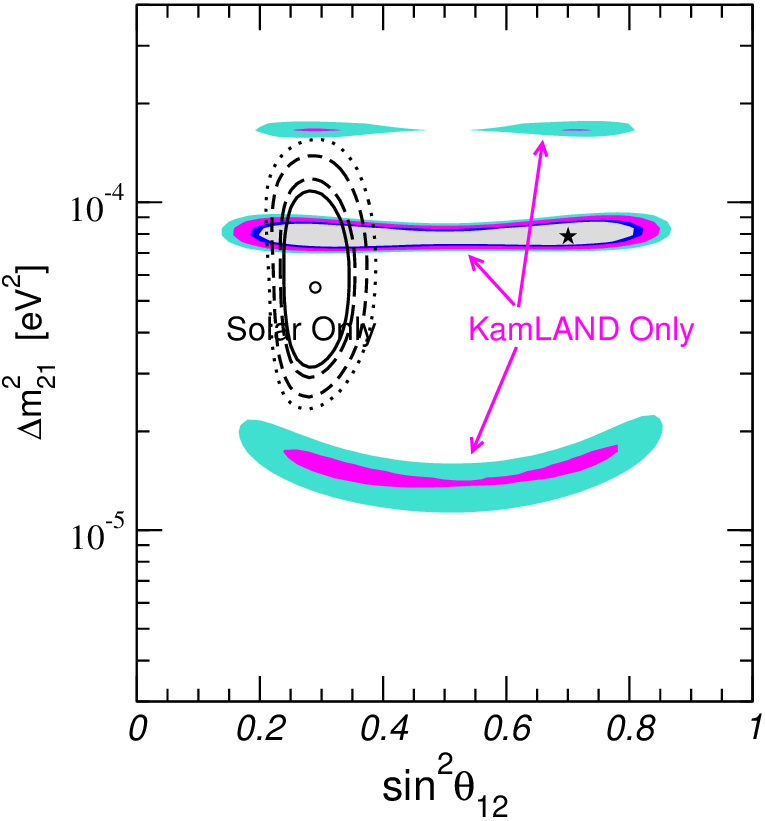}
&
\includegraphics[height=7.5cm]{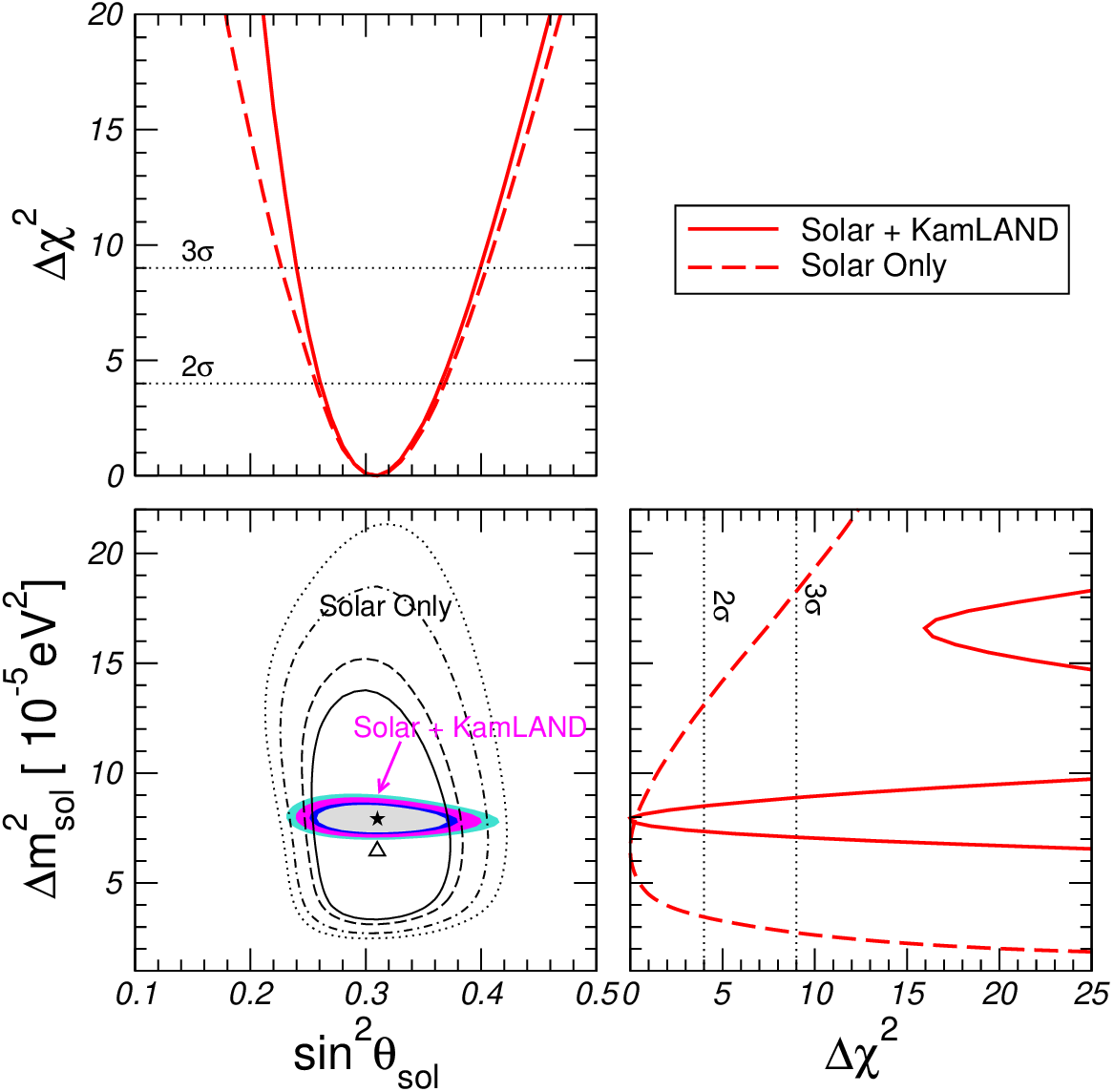}
\end{tabular}
\mycaption{Left: Allowed regions for $\sin^2\theta_{12}$ and $\Delta
  m^2_{21}$ at 90\%, 95\%, 99\%, and $3\sigma$ C.L.\ for solar
  neutrino data (contour lines) and the KamLAND reactor experiment
  (coloured regions). Right: Combined solar+KamLAND analysis. Also
  shown is the allowed region from solar data only (contour lines).}
  \label{fig:sol}
\end{figure}

{\it The 'solar parameters'.} In Fig.~\ref{fig:sol} the allowed
regions for $\theta_{12}$ and $\Delta m^2_{21}$ from analyses of solar
and KamLAND data are shown. Details of our solar neutrino analysis can
be found in Ref.~\cite{Maltoni:2003da} and references therein. We use
the same data as in Ref.~\cite{Maltoni:2004ei} from the Homestake,
SAGE, GNO, and Super-K experiments~\cite{solar}, and the SNO day-night
spectra from the pure D$_2$O phase~\cite{ahmad:2002ka}, but the CC,
NC, and ES rates from the SNO salt-phase are updated according to the
latest 2005 data~\cite{sno2005}. 
For the KamLAND analysis we are using the data equally binned in
$1/E_\mathrm{pr}$ ($E_\mathrm{pr}$ is the prompt energy deposited by
the positron), and we include earth matter effects and flux
uncertainties following Ref.~\cite{Huber:2004xh} (see the appendix of
Ref.~\cite{Maltoni:2004ei} for further details). We observe from the
figure a beautiful agreement of solar and KamLAND data. Moreover, the
complementarity of the two data sets allows a rather precise
determination of the oscillation parameters: The evidence of spectral
distortion in KamLAND data provides a strong constraint on $\Delta
m^2_{21}$, and leads to the remarkable precision of $4\%$ at $1\sigma$
(compare Tab.~\ref{tab:3nu-summary}). As visible in the right panel of
Fig.~\ref{fig:sol}, alternative solutions around $\Delta m^2_{21}\sim
2\times 10^{-4}$~eV$^2$ ($\sim 1.4\times 10^{-5}$~eV$^2$), which are
still present in the KamLAND-only analysis at 99\% C.L., are ruled out
from the combined KamLAND+solar analysis at about $4\sigma$
($5\sigma$). In contrast to $\Delta m^2_{21}$, the determination of
the mixing angle is dominated by solar data. Especially recent results
from the SNO experiment provide a strong upper bound on
$\sin^2\theta_{12}$, excluding maximal mixing at more than $5\sigma$.

\begin{figure}
\centering
\begin{tabular}{c@{\quad}c}
\includegraphics[height=6.5cm]{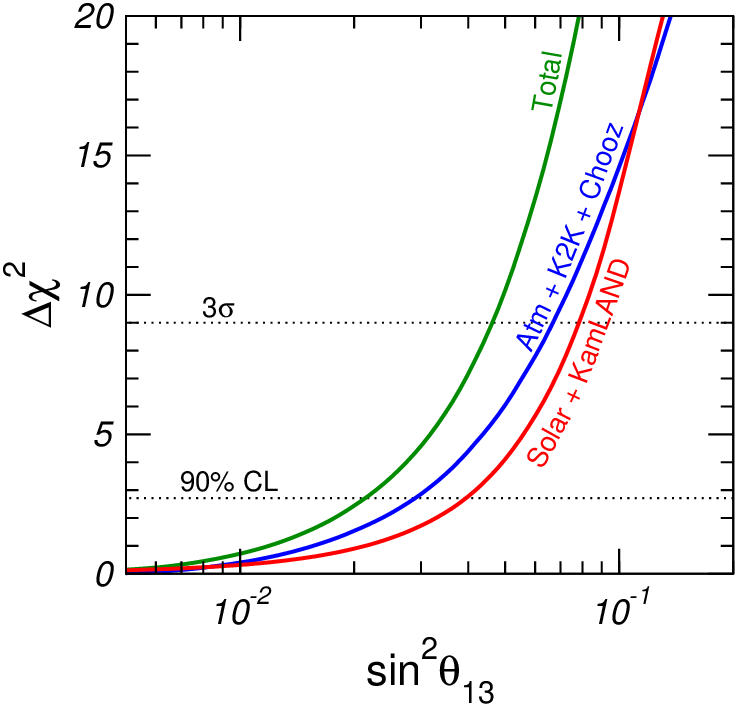}
&
\includegraphics[height=6.5cm]{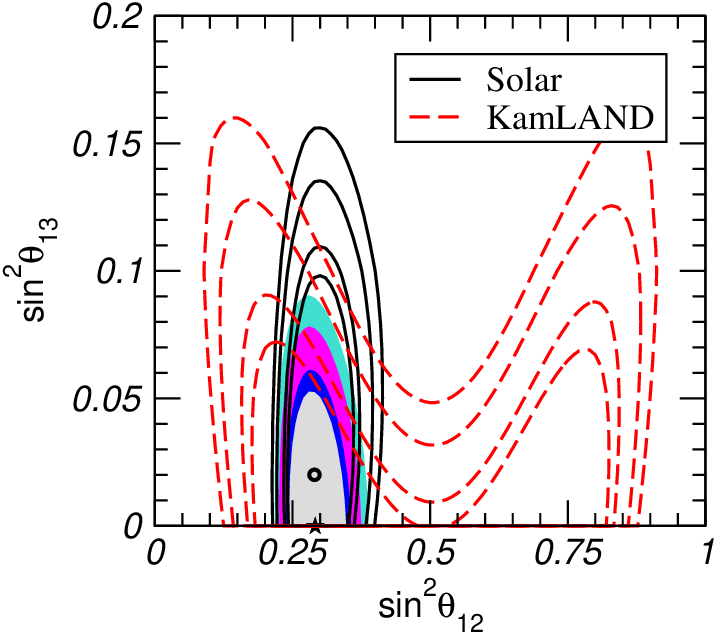}
\end{tabular}
\mycaption{Left: $\Delta\chi^2$ as a function of $\sin^2\theta_{13}$ for
   solar+KamLAND, CHOOZ+atmo{\-}spheric+K2K, and global data. Right:
   Allowed regions for $\sin^2\theta_{12}$ and $\sin^2\theta_{13}$ at
   90\%, 95\%, 99\%, and $3\sigma$ C.L.\ for solar neutrino data
   (solid contour lines), KamLAND (dashed contour lines), and
   solar+KamLAND data (coloured regions).}
   \label{fig:theta13}
\end{figure}

{\it The bound on $\theta_{13}$.} For the third mixing angle currently
only an upper bound exists. This bound is dominated by the CHOOZ
reactor experiment~\cite{chooz}, in combination with the $\Delta
m^2_{31}$-determination from atmospheric and K2K experiments. However,
recent improved data of solar and KamLAND experiments lead to a
non-negligible contribution of these experiments to the global bound,
especially for low values of $\Delta m^2_{31}$ within the present
allowed range~\cite{Maltoni:2003da}. From the left panel of
Fig.~\ref{fig:theta13} one deduces the following limits at 90\% C.L.\
($3\sigma$): 
\begin{equation}
  \sin^2\theta_{13} < \left\{
  \begin{array}{l@{\qquad}l}
     0.029 \:(0.067) & \mbox{CHOOZ+atm+K2K,}\\
     0.041 \:(0.079) & \mbox{solar+KamLAND,}\\
     0.021 \:(0.046) & \mbox{global data.}  
  \end{array}
  \right.
\end{equation}
In the right panel of Fig.~\ref{fig:theta13} we illustrate how the
combination of solar and KamLAND data leads to a non-trivial bound on
$\theta_{13}$. The allowed regions in the plane of $\sin^2\theta_{12}$
and $\sin^2\theta_{13}$ show the complementarity of the two data sets,
which follows from the very different conversion mechanisms: vacuum
oscillations at a baseline of order 180~km for KamLAND, and adiabatic
MSW conversion inside the sun for solar neutrinos. For a further
discussion see the appendix of Ref.~\cite{Maltoni:2004ei} or
Ref.~\cite{Goswami:2004cn}.


\section{Prospects for the coming ten years}

In this section I discuss the potential of long-baseline experiments,
from which results are to be expected within the coming ten years. The
first results will be obtained by the conventional beam experiments
\minos\ and the CNGS experiments \icarus\ and \opera. Subsequent
information might be available from new reactor experiments. We take
as examples \dchooz\ as a first stage experiment, and a generic
second-generation experiment labeled \RII, which could be realized at
sites in Brazil, China, Japan, Taiwan, or USA, see
Ref.~\cite{Anderson:2004pk} for an overview. Towards the end of the
anticipated time scale results from super-beam experiments \ttk\ and
\nova\ could be available.
The main characteristics of these experiments are summarized in
Tab.~\ref{tab:lbl}. For the simulation the GLoBES
software~\cite{globes} is used. Technical details and experiment
descriptions can be found in
Refs.~\cite{Huber:2003pm,Huber:2004ug,HLW}.
In the following I discuss the expected improvement on the leading
atmospheric parameters, and the sensitivity to $\theta_{13}$, the CP
phase $\deltacp$ and the type of the neutrino mass hierarchy. A
discussion of prospects to improve the determination of the leading
solar parameters can be found in Ref.~\cite{solar-params}.

\begin{table}
\centering
\begin{tabular}{|lrrrl|}
\hline\hline
 Label & $L$ & $\langle E_\nu \rangle$ & $t_{\mathrm{run}}$ & channel \\
\hline
\multicolumn{5}{|l|}{\bf{Conventional beam experiments:}}\\
 \minos~\cite{minos} & $735 \, \mathrm{km}$ & $3 \,\mathrm{GeV}$ & $5 \, \mathrm{yr}$ & 
 $\nu_\mu \to \nu_\mu,\nu_e$\\
 \icarus~\cite{icarus} &  $732\,\mathrm{km}$ &  $17\,\mathrm{GeV}$ & $5 \, \mathrm{yr}$ & 
 $\nu_\mu \to \nu_e, \nu_\mu, \nu_\tau$\\
 \opera~\cite{opera} &   $732\,\mathrm{km}$ &  $17\,\mathrm{GeV}$ & $5 \, \mathrm{yr}$ & 
 $\nu_\mu \to \nu_e, \nu_\mu, \nu_\tau$\\
\multicolumn{5}{|l|}{\bf{Reactor experiments with near and far detectors:}}\\
 \dchooz~\cite{Ardellier:2004ui} & $1.05 \, \mathrm{km}$ & $\sim 4 \, \mathrm{MeV}$ & $3 \, \mathrm{yr}$ & 
 $\bar\nu_e \to \bar\nu_e$\\
 \RII~\cite{Anderson:2004pk} & $1.70 \, \mathrm{km}$ & $\sim 4 \, \mathrm{MeV}$ & $5\,\mathrm{yr}$ & 
 $\bar\nu_e \to \bar\nu_e$\\
\multicolumn{5}{|l|}{\bf{Off-axis super-beams:}} \\
 \ttk~\cite{Itow:2001ee} &  $295  \, \mathrm{km}$ & $0.76 \, \mathrm{GeV}$ & $5 \, \mathrm{yr}$ & 
 $\nu_\mu \to \nu_e,\nu_\mu$\\
 \nova~\cite{Ayres:2004js} & $812 \, \mathrm{km}$ & $2.22 \, \mathrm{GeV}$ & $5 \, \mathrm{yr}$ & 
 $\nu_\mu \to \nu_e,\nu_\mu$\\
\hline\hline
\end{tabular}
\mycaption{Summary of upcoming experiments.}
\label{tab:lbl}
\end{table}

\begin{figure}
   \centering
   \includegraphics[width=0.8\textwidth]{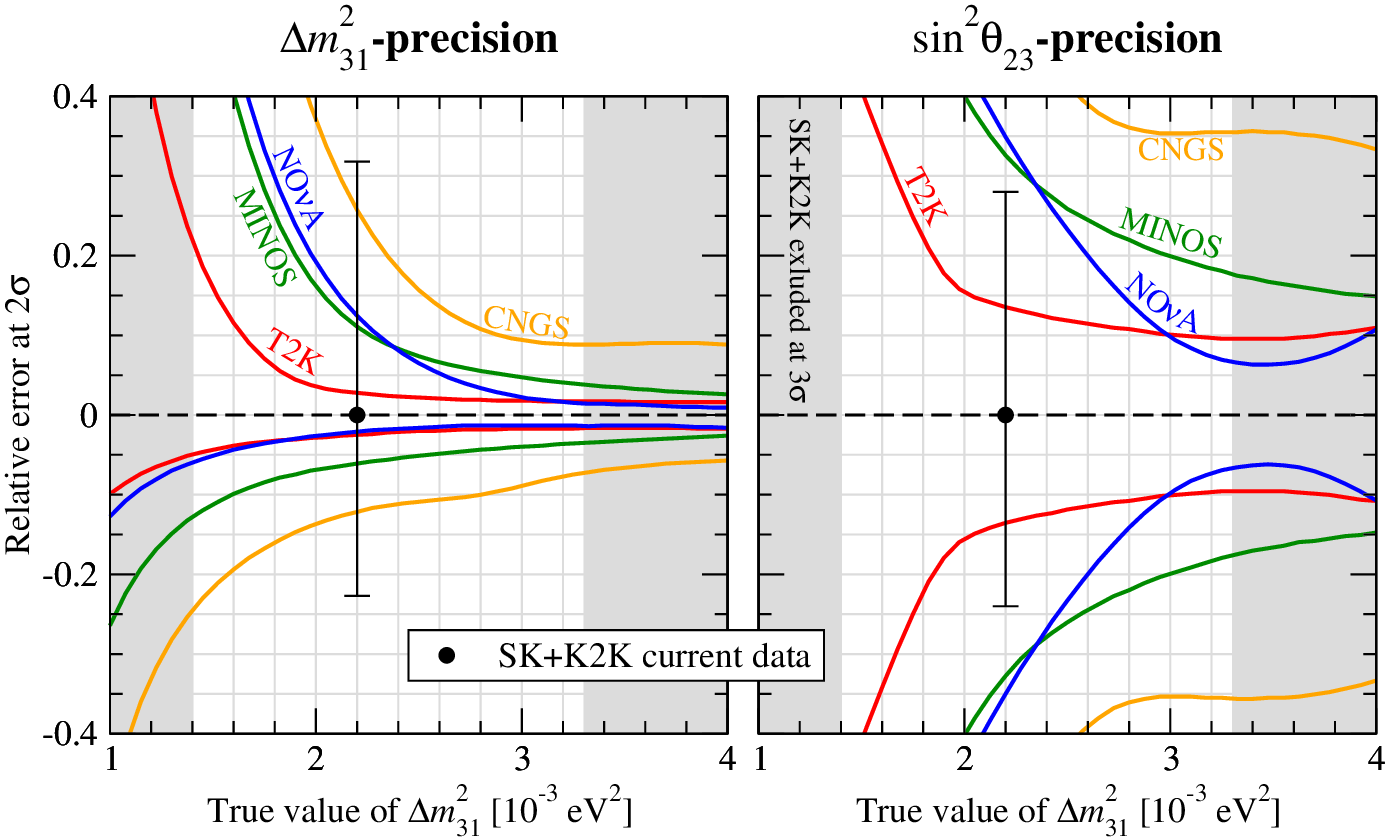}
   \mycaption{Prospective relative errors at $2\sigma$ on $|\Delta
   m^2_{31}|$ (left) and $\sin^2\theta_{23}$ (right) as a function of
   the true value of $\Delta m^2_{31}$ and for the true value
   $\sin^2\theta_{23} = 0.5$. The dots with the error bars indicate
   the present accuracy at $2\sigma$ from atmospheric and K2K data.
   The gray shaded region is excluded at $3\sigma$ by present data.}
   \label{fig:atm-prospect}
\end{figure}

{\it The 'atmospheric parameters'.} In Fig.~\ref{fig:atm-prospect} I
show how the upcoming experiments will improve the accuracy on
$|\Delta m^2_{31}|$ and $\sin^2\theta_{23}$. Already \minos\ will
decrease significantly the error on $|\Delta m^2_{31}|$. The
non-trivial constraint on $|\Delta m^2_{31}|$ from the CNGS
experiments results from the fact that in this analysis also the
$\nu_\mu$ disappearance channel is included for these experiments (see
Ref.~\cite{Huber:2004ug} for details). For not too small values of
$|\Delta m^2_{31}|$ \ttk\ will provide a determination of order 2\% at
$2\sigma$, which corresponds roughly to an improvement of one order of
magnitude with respect to the present error. However, from the right
panel of Fig.~\ref{fig:atm-prospect} one observes that for most values
of $|\Delta m^2_{31}|$ \ttk\ will improve the accuracy on
$\sin^2\theta_{23}$ only by a factor of 2 with respect to the present
uncertainty. The main reason for this only modest improvement comes
from the fact that disappearance experiments measure
$\sin^22\theta_{23}$, and a small uncertainty on $\sin^22\theta_{23}$
translates into a relatively large error for $\sin^2\theta_{23}$, if
$\theta_{23}$ is close to maximal mixing.

It is interesting to note that the lower bound on $|\Delta m^2_{31}|$
from \nova\ is comparable to the one from \ttk, however, the upper
bound is significantly weaker because of a strong correlation between
$|\Delta m^2_{31}|$ and $\sin^2\theta_{23}$. Also the
$\sin^2\theta_{23}$ measurement of \nova\ is affected by this
correlation, which gets resolved only in the range $|\Delta m^2_{31}|
\gtrsim 3\times 10^{-3}$~eV$^2$. This correlation appears because the
\nova\ detector is optimized for electrons, whereas the atmospheric
parameters are determined essentially by the $\nu_\mu$ disappearance
channel. Let me add that as in Ref.~\cite{Huber:2004ug} we assume here
a low-Z-calorimeter. Using the totally active scintillator detector
(TASD) as proposed in Ref.~\cite{Ayres:2004js} improves the
performance of \nova\ for the atmospheric parameters.

\begin{figure}
   \centering \includegraphics[width=0.6\textwidth]{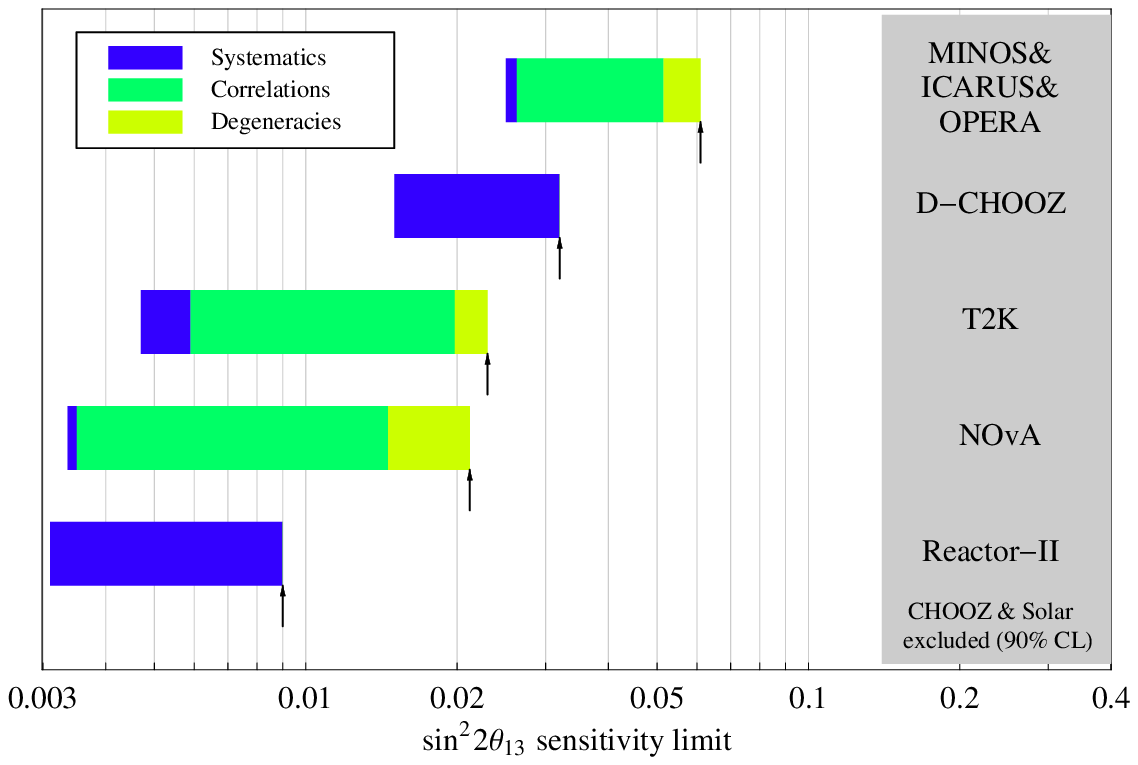}
   \mycaption{Sensitivity to $\sin^22\theta_{13}$ at 90\% C.L. The left
   edges of the bars are obtained for the statistics limits only,
   whereas the right edges are obtained after successively switching
   on systematics, correlations, and degeneracies, {\it i.e.}, they
   correspond to the final sensitivity.  The gray shaded region is
   excluded at 90\%~C.L.\ by present data. For the true values of the
   oscillation parameters, we use $\Delta m_{31}^2 = +2.0 \times
   10^{-3}\,\mathrm{eV}^2$, $\sin^22\theta_{23} = 1$, $\Delta m_{21}^2
   = 7.0 \times 10^{-5}\,\mathrm{eV}^2$, $\sin^22\theta_{12} = 0.8$.}
   \label{fig:th13}
\end{figure}

{\it The limit on $\theta_{13}$.} Fig.~\ref{fig:th13} shows the limits
which can be obtained on $\stheta$ by the experiments under
consideration, if no signal for a finite $\theta_{13}$ is found. One
observes that the conventional beams may improve the present limit
roughly by factor of 2, \dchooz\ by a factor of 4, and the super-beams
by a factor of 6. An optimized reactor experiment could improve the
present bound by more than one order of magnitude, reaching a limit
for $\stheta$ below $10^{-2}$ at 90\% C.L. The limits shown in
Fig.~\ref{fig:th13} depend on the true value of $\Delta m^2_{31}$; in
general stronger limits can be reached for larger values of $\Delta
m^2_{31}$.

The figure illustrates that the $\stheta$ limits from
$\nu_\mu\to\nu_e$ appearance beam experiments are strongly affected by
correlations, mainly between $\stheta$ and $\deltacp$. The bars
labeled `Degeneracies' originate from the fact that the hierarchy
cannot be determined, and the sensitivity is quoted for the hierarchy
which gives the worse limit (see the appendix of
Ref.~\cite{Huber:2004ug} for a detailed discussion of the sensitivity
limit). Clearly, the $\stheta$ limit from reactor experiments is
completely free from correlations and
degeneracies~\cite{Huber:2003pm,minakata}, since the $\bar\nu_e$
survival probability does not depend on $\deltacp$ and 
$\theta_{23}$, and the dependence on the solar parameters is
negligibly small. These experiments are dominated by systematical
uncertainties, related mainly to the comparison of the near and far
detectors. A possibility to reduce significantly the impact of these
uncertainties has been presented in Ref.~\cite{Huber:2004bh}.

\begin{figure}
\centering
   \includegraphics[width=\textwidth]{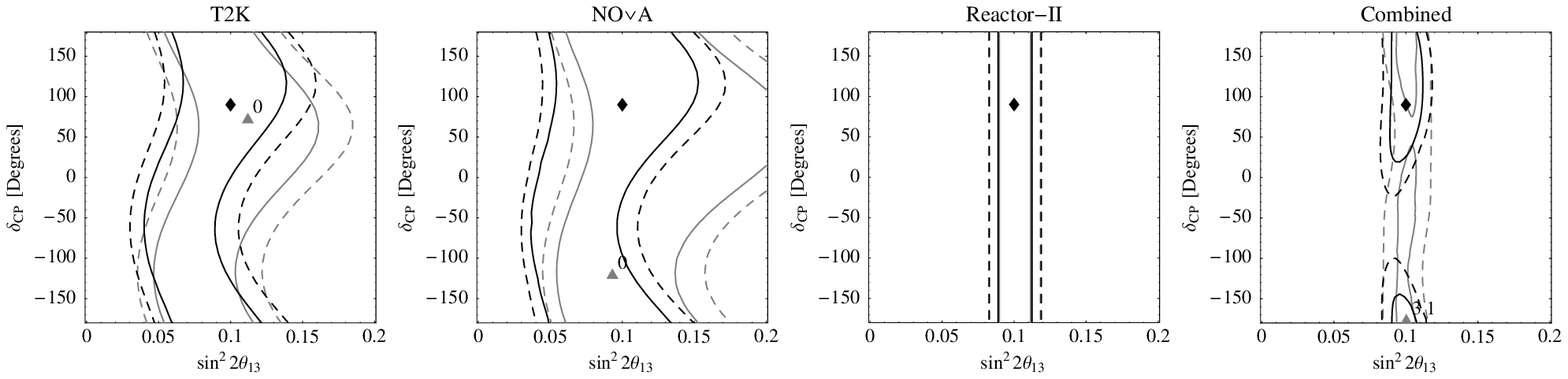}
   \mycaption{\label{fig:deltatheta1} The $90 \%$ C.L.\ (solid) and $3
   \sigma$ (dashed) allowed regions (2 d.o.f.) in the
   $\sin^22\theta_{13}$-$\deltacp$ plane for the true values
   $\stheta=0.1$ and $\deltacp=90^\circ$. The black curves refer to
   the allowed regions for the normal mass hierarchy (assumed to be
   the true hierarchy), whereas the gray curves refer to the
   $\mathrm{sgn}(\Delta m^2_{31})$-degenerate solution (inverted
   hierarchy), where the projections of the minima onto the
   $\stheta$-$\deltacp$ plane are shown as diamonds (normal hierarchy)
   and triangles (inverted hierarchy). For the latter, the
   $\Delta\chi^2$-value with respect to the best-fit point is also
   given.}
\end{figure}

{\it Possibilities for large $\theta_{13}$.} Finally I discuss the
potential of the experiments of Tab.~\ref{tab:lbl} if $\theta_{13}$ is
not too far from the present upper bound. In
Fig.~\ref{fig:deltatheta1} the allowed regions in the
$\stheta$-$\deltacp$ plane are shown assuming a true value of $\stheta
= 0.1$. For the super-beam experiments \ttk\ and \nova\ one observes
the strong correlation between $\stheta$ and $\deltacp$, which
introduces a large uncertainty in $\stheta$ but does not permit to
draw any conclusions on $\deltacp$. In contrast, the reactor
experiment provides an accurate measurement of $\stheta$ with an
accuracy at the level of 10\% at 90\% C.L.
None of the experiments on its own can identify the mass hierarchy. As
indicated in the figure, the best fit point of the wrong hierarchy is
as good as the true best fit point. However, from the combination of
\ttk+\nova+\RII\ the wrong hierarchy can be disfavoured with
$\Delta\chi^2 = 3.1$, which corresponds roughly to the 90\% C.L.  The
crucial element for the mass hierarchy determination is the long
baseline of 812~km for \nova, leading to matter effects which allow to
distinguish between normal and inverted hierarchy. Let me add, that
the hierarchy sensitivity strongly depends on the true value of
$\deltacp$; typically it is optimal for $\deltacp \simeq -90^\circ$, and
worst for $\deltacp\simeq 90^\circ$~\cite{Huber:2004ug}.

As visible from Fig.~\ref{fig:deltatheta1} no information can be
obtained on CP violation, {\it i.e.}, at least one of the
CP-conserving values $\deltacp=0,180^\circ$ is contained within the
90\% C.L.\ region, even though the assumed true value
$\deltacp=90^\circ$ corresponds to maximal CP violation. The reason is
that no anti-neutrino data is included in this analysis, since due to
the low cross sections it seems unlikely that significant anti-neutrino
data will be available within the anticipated time scale. Note
however, that from the combined analysis some values of $\deltacp$ can
be excluded for a given mass hierarchy.


\section{Conclusions}

I have reviewed the present status of neutrino oscillations from
world neutrino oscillation data, including solar, atmospheric, reactor
and accelerator experiments. The results of a global analysis within
the three-flavour framework have been presented, and in particular, the
bound on $\theta_{13}$, which emerges from the interplay of various
data sets has been discussed. Furthermore, a prospect on where we
could stand in neutrino oscillations in ten years from now has been given.
Based on a simulation of upcoming long-baseline accelerator and
reactor experiments the improvements on the leading atmospheric
parameters, as well as the sensitivity to $\theta_{13}$, $\deltacp$
and the neutrino mass hierarchy have been discussed.

\subsection*{Acknowledgments} 

I thank the organizers for the very pleasant and interesting
conference.  The results presented here have been obtained in
collaboration with P.~Huber, M.~Lindner, M.~Maltoni, M.~Rolinec,
M.A.~T{\'o}rtola, J.W.F.~Valle and W.~Winter. T.S.\ is supported by a
``Marie Curie Intra-European Fellowship within the 6th European
Community Framework Program.''

\end{document}